\documentclass[12pt,manuscript]{aastex}

\usepackage{txfonts}

\usepackage{graphicx}

\usepackage{longtable}

\usepackage{natbib}

\slugcomment{For submission to the Astrophysical Journal}
\shorttitle{Repeating Chromospheric Nanoflares}
\shortauthors{Bradshaw and Klimchuk}

\begin{document}
\tighten
\title{Chromospheric Nanoflares as a Source of Coronal Plasma: II. Repeating Nanoflares}

\author{S. J. Bradshaw}
\affil{Department of Physics and Astronomy, Rice University, Houston, TX 77005, USA}
\email{stephen.bradshaw@rice.edu}\and

\author{J. A. Klimchuk}
\affil{NASA Goddard Space Flight Center, Greenbelt, MD 20771, USA}
\email{James.A.Klimchuk@nasa.gov}


\begin{abstract}
The million degree plasma of the solar corona must be supplied by the underlying layers of the atmosphere. The mechanism and location of energy release, and the precise source of coronal plasma, remain unresolved. In earlier work we pursued the idea that warm plasma is supplied to the corona via direct heating of the chromosphere by nanoflares, contrary to the prevailing belief that the corona is heated in-situ and the chromosphere is subsequently energized and ablated by thermal conduction. We found that single (low-frequency) chromospheric nanoflares could not explain the observed intensities, Doppler-shifts, and red/blue asymmetries in Fe XII and XIV emission lines. In the present work we follow up on another suggestion that the corona could be powered by chromospheric nanoflares that repeat on a timescale substantially shorter than the cooling/draining timescale. That is, a single magnetic strand is re-supplied with coronal plasma before the existing plasma has time to cool and drain. We perform a series of hydrodynamic experiments and predict the Fe XII and XIV line intensities, Doppler-shifts, and red/blue asymmetries. We find that our predicted quantities disagree dramatically with observations and fully developed loop structures cannot be created by intermediate- or high-frequency chromospheric nanoflares. We conclude that the mechanism ultimately responsible for producing coronal plasma operates above the chromosphere, but this does not preclude the possibility of a similar mechanism powering the chromosphere; extreme examples of which may be responsible for heating chromospheric plasma to transition region temperatures (e.g. type~II spicules).
\end{abstract}



\section{Introduction}
\label{introduction}

The hot ($T>10^6$~K) plasma observed in the solar corona must originate from the lower-lying layers of the solar atmosphere. In particular, the chromosphere ($T\sim10^4$~K) is thought to serve as a reservoir of coronal material. The problem yet to be solved is to determine the physical mechanism by which chromospheric material is heated and lifted to form the corona. In the prevailing scenario, free energy is released from the magnetic fields threading the corona and then locally dissipated in the plasma. This serves to heat low-density plasma, in-situ, to extremely high temperatures ($T>10^6$~K). The radiative losses from the corona are not sufficient to remove the excess energy and strong thermal conduction fronts develop to carry it away into the lower atmosphere. The upper chromosphere is strongly heated and the resulting pressure imbalance causes it to expand, filling the corona with hot plasma and raising its density. If the coronal heating were to remain steady then a hydrostatic equilibrium would be reached in which the temperature and density evolve to constant values, and the bulk flows fall to zero; the energy input is then balanced by the radiative losses and downward thermal conduction. In the case of impulsive heating the plasma cools and drains from the corona when the energy input is switched off.  \citep[These scenarios are discussed in detail in~][]{Klimchuk2006,Klimchuk2008,Cargill2012,Reale2014}.

It is difficult to reconcile steady heating with observations of the solar corona, for example: the measured over-densities at 1~MK \citep{Aschwanden2000} can be explained by impulsive heating models \citep{Spadaro2003}; the emission measure {\it EM} distribution, quantified by the observationally measured slope between the temperature of peak {\it EM} and 1~MK, is consistent with impulsive heating \citep{Reep2013,Cargill2014}; and the faint glimpses of super-heated ($T\sim10$~MK) plasma in non-flaring active regions  \citep{McTiernan2009,Schmelz2009a,Schmelz2009b,Reale2009a,Reale2009b,Testa2011,Miceli2012,Testa2012,Brosius2014,Petralia2014a,Caspi2015} have long been considered to be the ``smoking gun'' of impulsive heating \citep{Cargill1994,Bradshaw2006,Reale2008}. Furthermore, we know of no plausible plasma process that can sustain a steady release of energy over an extended period of time ($\sim$~hours). The evidence for impulsive heating powering the corona is gradually mounting, but there is much that we do not yet know for certain about the underlying mechanism: where does it operate and over what spatial scale? How much energy is released per heating event? What are the timescales and frequencies of the events? What is the nature of the mechanism itself? Impulsive heating events are generally referred to as {\it nanoflares}. We will use this term for consistency with the published literature, but no specific heating mechanism is implied. For example, the free magnetic energy may be impulsively released and dissipated in the corona by small-scale magnetic reconnection events \citep{Parker1988} or via Alf\'{e}n wave damping \citep{Asgari-Targhi2013,Ofman1998}.

An alternative possibility to coronal heating is that the release and dissipation of magnetic energy actually takes place in the chromosphere. In this scenario, the chromospheric plasma is directly heated to coronal temperatures, rather than by thermal conduction fronts that are driven into the lower atmosphere as a consequence of heating localized in the corona \citep{Hansteen2010}. One manifestation of this process might be the type~II spicules \citep{DePontieu2007} that have been suggested to play a role in supplying mass and energy to the corona \citep{DePontieu2009,DePontieu2011,Moore2011}. \cite{Raouafi2015} have argued that type~II spicules have much more in common with so-called ``classical'' spicules \citep{Beckers1968,Beckers1972,Pasachoff2009} than do type~I spicules. In particular, type~II and classical spicules are both commonly observed in quiet Sun and coronal hole regions, whereas type~I spicules appear to be exclusively confined to active regions. Furthermore, \cite{Pereira2013} artificially coarsened Hinode/SOT observations of type~II spicules to demonstrate that their lifetimes and ejection speeds are consistent with the earlier, ground-based observations of classical spicules ($\sim5$~minutes and $25$~km~s$^{-1}$). \citep{Raouafi2015} suggest that the term ``classical spicules'' be used for the earlier objects \citep[see][for a review]{Sterling2000} and the terms type~I and II spicules be reserved for spicular phenomena observed during the era of Hinode observations. The key properties of type~II spicules are their faster velocities ($30-110$~km~s$^{-1}$) and shorter lifetimes ($50-150$~seconds) compared with type~I spicules. Type II spicules are typically observed in Ca~II emission by SOT before fading out of that passband and appearing in the SDO/AIA 304~\AA~channel, indicating that some degree of heating takes place as they rise. There is some observational evidence for a transition region or coronal component of their emission, visible as a bright, moving front in the AIA~171~\AA~channel \citep[e.g.][]{DePontieu2011}. The two possibilities for producing this warmer emission are: (1) pre-existing coronal material is shock-heated as the upflowing spicular material rams into it \citep{Klimchuk2012,Petralia2014b}; or (2) the tip of the spicule is heated by some in-situ process, which must be impulsive because the cooler emission quickly disappears from view \citep{DePontieu2007}.

\cite{Klimchuk2012} investigated the suggestion that type~II spicules with impulsively heated tips can supply the corona with most of its hot plasma. He showed that three quantities are predicted by this scenario, which can be compared with observational measurements: (i) the ratio of the blue-wing intensity to the line core intensity; (ii) the ratio of the transition region {\it EM} ($\leq 10^5$~K) to the coronal {\it EM}; and (iii) the ratio of the density measured from the blue-wing emission to the line core density for density-sensitive emission lines. The predicted ratios differed from those observed by several orders of magnitude \citep{Patsourakos2014,Tripathi2013} and it was concluded that spicules could account for only a small fraction of the hot coronal plasma. We followed up the analytical work of \cite{Klimchuk2012} with a detailed numerical study in \cite{Klimchuk2014} (henceforth referred to as KB14). This extended the earlier work because the generality of our numerical approach allowed chromospheric nanoflares to be treated as a general phenomenon, not required to be associated with spicules. Energy was injected into the upper 1000~km of the chromosphere, which did not lead to an ejection of cool material in any of our experiments. The model set-up was nonetheless relevant to case (2) above for producing warm emission because the dynamics of the spicule tip is determined by explosive expansion, just as though it were directly heated in the chromosphere. Our findings supported those of \cite{Klimchuk2012}. We predicted total intensities for coronal lines (from Fe~XII and XIV) much fainter than those observed, the blue-shifted velocities were too fast, the red/blue asymmetries were too large, and any significant warm emission was confined to much lower altitudes than where it is typically observed. In consequence, we concuded that chromospheric nanoflares are not the main source of coronal plasma, but may play a role in powering the chromosphere and/or generating waves that may contribute to coronal heating through some other mechanism.

Our previous findings are based on single heating events occurring along isolated magnetic strands, where each strand ultimately returns to its cool and tenuous initial state between events. The question that we address in the present work asks what happens if the strand does not substantially cool and drain before the next heating event? Motivating this study is the suggestion of \cite{DePontieu2011} that spicules may occur much more frequenty, based on H$\alpha$ observations of rapid blue-shift events (RBEs), with recurrence times as short as 500~s at the same location. The warm material must gradually build up to sufficiently high densities in the corona in this scenario, and the mass loss through draining is then offset by continued ejections. \cite{Klimchuk2012} identified two difficulties with this idea in terms of explaining the warm corona. Firstly, if each spicule contributes a few percent of the strand's coronal material then $\sim20$ spicules are required (for a conservative supply of 5\% per spicule, assuming weak radiation and no draining). At the highest recurrence frequency observed, of one spicule every 500~s,  then a strand would take $\sim10^4$~s to fully evolve. However, loops in the range $1-2$~MK have total lifetimes (complete brightening and fading cycle) of $1000-5000$~s \citep{Klimchuk2010}. Secondly, maximum temperatures are observed close to the strand apex, but each spicule energizes only the lower parts of each strand. An upward thermal conduction flux is then needed to take over the energy transport and power the corona, which requires the temperature maximum to be located close to the foot-point, but this is inconsistent with the observed temperature structure.

The work reported here investigates whether high-frequency chromospheric nanoflares can lead to an accumulation of warm material and produce fully developed coronal strands with the observed properties. As in our previous studies we forward model the profiles of strong coronal lines (Fe~XII at 195.119~\AA~and Fe~IV at 274.204~\AA) to predict their intensities, Doppler-shifts, and red/blue asymmetries, to compare with real observations. In Section~\ref{model} we give the details of our numerical model and experiments, in Section~\ref{results} we present our findings, and in Section~\ref{SandC} we provide a summary and present our conclusions.

\section{Numerical Model}
\label{model}

\begin{figure}
\includegraphics[width=0.5\textwidth]{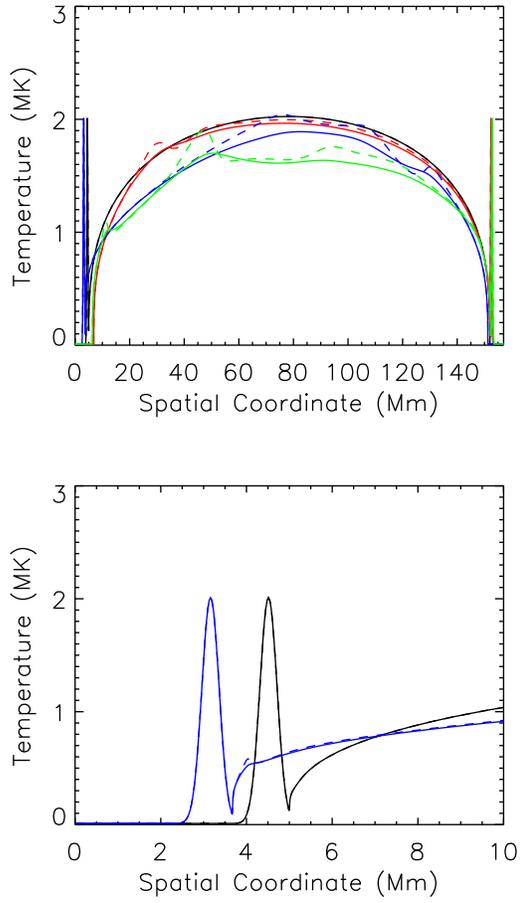}
\caption{Upper panel: the temperature profiles of the first four nanoflares during Run~1 at 0~(black), 250~(red), 500~(blue), and 750~s~(green). Lower panel: the temperature profiles at the left-hand footpoint at 0 and 500~s. The solid line indicates the electron temperature and the dashed line indicates the hydrogen temperature.}
\label{fig1}
\end{figure}

Our modeling process follows the approach of KB14. The interaction between the magnetic field and the plasma doubtless drives the energy release during chromospheric nanoflares or other events \citep[e.g.][]{Sykora2013} in some manner, though the precise nature of the mechanism has not yet been elucidated. Nonetheless, once the energy has been released and the temperature reaches millions of degrees Kelvin then the plasma becomes fully ionized (except for heavy trace elements), strongly confined to the magnetic strand with negligible cross-field energy transport \citep[$\kappa_{\perp}/\kappa_{\parallel} \sim 10^{-8}$;][]{Goedblood2004}, and its evolution consequently dominated by field-aligned processes. There may be a small amount of lateral expansion of the heated flux tube, but given the thermal pressures encountered in our numerical experiments and the magnetic pressure in active regions then we expect the $\beta$ parameter to remain relatively low $(\sim 10^{-3})$ and the plasma motion to be primarily directed upward into lower pressure regions of the corona. The field-aligned evolution of the plasma, subject to repeating chromospheric nanoflares, is modeled using the HYDRAD code \citep{Bradshaw2013}. The key capabilities of HYDRAD that make it particularly well-suited to this investigation are its ability to: calculate the time-dependent ionization state of the plasma; and fully resolve multiple regions of steep gradients by adaptive mesh refinement (AMR). Non-equilibrium ionization (NEI) effects can be important due to the fast temperature changes that take place when the plasma is strongly heated or as it rapidly cools under expansion. Furthermore, the strong flows predicted by the model can quickly transport ions across steep gradients into temperature regimes outside their formation temperature range in equilibrium. Additional transition regions can form along the magnetic strand as chromospheric nanoflares drive the plasma up to coronal temperatures (Figure~\ref{fig1}) and AMR is essential to properly resolve each of them in order to accurately capture the dynamic behavior \citep{Bradshaw2013}. The code refines down to a cell size of 3.9~km in the present work, which we have determined is ample to support the heat flux from a $2-3$~MK corona at transition region temperatures.

In common with KB14 we model a coronal loop as a semi-circular magnetic strand of length 157~Mm, yielding an apex height of 50~Mm \citep[also consistent with][]{Klimchuk2012}. The choice of loop length is relatively unimportant since most of the evolution that determines the predicted observations happens at very low altitudes: sudden heating of the upper chromosphere, followed by rapid expansion and cooling. The lower 5~Mm of each leg comprise a gravitationally stratified chromosphere with a uniform temperature of 0.01~MK. The chromosphere thus spans multiple density scale heights such that the evolution of the overlying plasma remains well isolated from the surface boundary conditions. The temperature is maintained at $\approx0.01$~MK by sharply decreasing the optically-thin radiative losses to zero over a narrow temperature range above 0.01~MK \citep[e.g.][]{Klimchuk1987}. While this method does not reproduce the expected chromospheric radiative losses, we are primarily interested in the dynamical evolution of the chromospheric plasma once it has been very rapidly heated to coronal temperatures where the radiative losses are treated correctly. Hence, our model chromosphere has the necessary physical properties, essentially as a reservoir/sink for coronal mass, to satisfy the requirements of this investigation.

We subject the loop to a nanoflare by heating the upper 1~Mm (1000~km) of the chromosphere. Sufficient thermal energy, with a Gaussian spatial distribution, is added to the plasma (at $t=0$~s, for example) such that the electron and ion temperature profiles both peak at 2~MK in the upper chromosphere. In the case of the nanoflare at $t=0$~s in Figure~\ref{fig1} the total energy required was $1.64\times10^{10}$~erg~cm$^{-2}$, and for the nanoflare at $t=500$~s the total energy required was $1.75\times10^{11}$~erg~cm$^{-2}$. The energy difference, as is clear from the lower panel of the Figure, arises because the later nanoflare takes place deeper in the atmosphere where the density is higher and therefore substantially more energy is required for the energy per particle to be consistent with a 2~MK plasma. Directly adding the energy instantaneously raises the temperature, but drives the plasma similarly to the 10~s nanoflares we explored in KB14. We chose to add energy directly, rather than via a parameterized heating function added to the energy equation (as is more typical), for reasons of efficiency when modeling repeated nanoflares. We have already seen that different amounts of energy are required to reach 2~MK for subsequent nanoflares as the transition region moves up and down the loop in response to the previous event. Since the rate of temperature change for rapid heating scales as $n$ and the rate of energy loss by radiation scales as $n^2$, estimating (by trial and error) the energy input to a heating function to guarantee 2~MK plasma for each nanoflare necessitated direct intervention with the model run, and proved time-consuming and cumbersome. Adding the energy directly, under the assumption of essentially instantaneous heating, allowed the modeling process to become more-or-less automated. 

As an aside, it is interesting to speculate here that the movement of the transition region as it responds to changes in the coronal pressure may lead to a natural form of intermittency in the temperature evolution. At relatively low coronal pressure the transition region resides at a higher altitude, allowing the chromosphere to expand and thereby also lowering the density of the upper chromosphere where the heating takes place. A heating event can then efficiently raise the plasma temperature. Conversely, at high coronal pressure the transition region is forced to a lower altitude, compressing the chromosphere and increasing the density of the upper chromosphere. An identical heating event may then be insufficient to appreciably raise the temperature because less energy is deposited per particle. This pressure dependent variability in density provides a neat regulation mechanism: only when the loop cools and drains beyond a certain point can it be reheated to high temperatures. The frequency at which hot plasma is produced is less than the frequency of nanoflares, since some nanoflares are unable to overcome the high densities. Whether nanoflares are effectively in the high-frequency or low-frequency regime will depend on the distribution of energies, delay between successive events, and other factors. However, this is beyond the scope of the present work since we aim to determine whether chromospheric nanoflares at any physically reasonable frequency (e.g. guided by observations) can sustain a coronal loop.

\begin{longtable}{c c c c}
\caption{Summary of the Key Parameters for the Numerical Experiments.}\\
\hline
Run~\# & Inter-Event Period & Duration & \# of Nanoflares\\
 & [s] & [s] & \\
\hline
\endhead
 1 & 250 & 10,000 & 40 \\
 2 & 500 & 10,000 & 20 \\
 3 & 1000 & 20,000 & 20 \\
\hline
\label{table1}
\end{longtable}

\begin{figure}
\includegraphics[width=\textwidth]{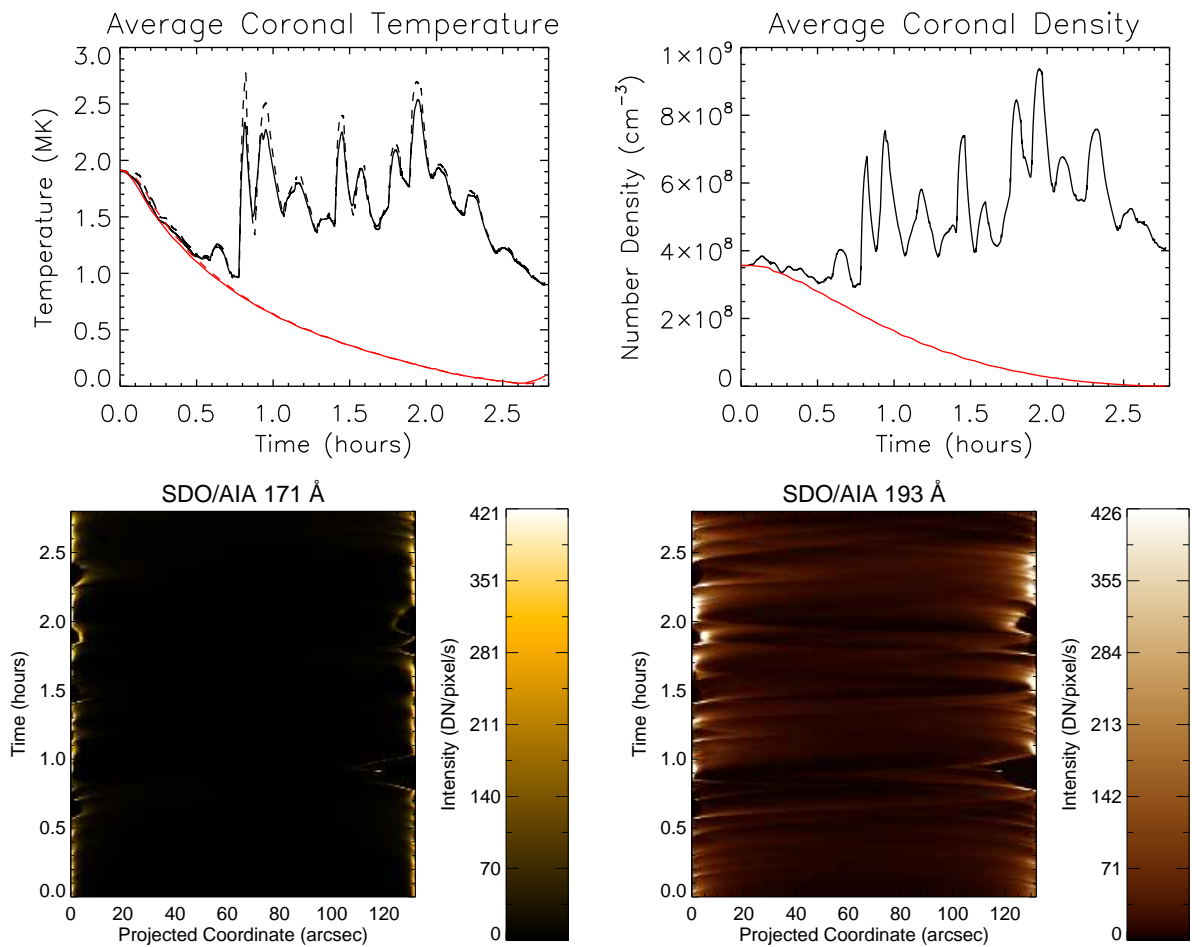}
\caption{Upper panels: the temperature (electron~[solid]; ion~[dashed]) and density, spatially averaged over the upper 50\% of the loop, as a function of time for Run~1; the red curves show the temperature and density for a corresponding run in which the plasma was simply allowed to cool. Lower panels: the predicted counts, as a function of position along the projected loop and time, forward modeled for the SDO/AIA 171 and 193~\AA~channels; the counts are integrated over 10~s intervals to $\sim$ emulate the cadence of AIA.}
\label{fig2}
\end{figure}

We impose nanoflares alternately at each footpoint with inter-event periods of 250, 500, and 1000~seconds (Runs 1 to 3, respectively). The inter-event periods at the same footpoint are then 500, 1000, and 2000~seconds. These periods were chosen based on the results of \cite{Cargill2014}, who found that event periods between a few hundred seconds and of order 1000 seconds could recover the observed range of coronal emission measure slopes (the slope between the temperature of peak {\it EM} and the {\it EM} at 1~MK). We also note that \cite{Ugarte-Urra2014} estimated a period of $\sim1400$~s from intensity fluctuations of the Fe~XVIII component in SDO/AIA~94~\AA~observations of active regions, and \cite{Dahlburg2005} estimated 2000~s based on simulations of the secondary instability. \cite{Klimchuk2015} argued that reconnection events should recur with an average period of $\sim200$~s, but noted that multiple events can combine so that the effective nanoflare period is longer. Each nanoflare heats the upper 1~Mm of the chromosphere to 2~MK, which is the characteristic coronal temperature that we need to explain. The upper panel of Figure~\ref{fig1} shows the temperature profiles of the first four nanoflares for Run~1 at 0, 250, 500, and 750~s. The lower panel shows just the left-hand footpoint and the temperature profiles at 0 and 500~s. 

Since we are primarily interested in determining whether repeating chromospheric nanoflares can support a hot, overlying corona, in a manner that is consistent with observations, then sufficient background heating is applied to produce a static equilibrium initial condition of apex temperature and density 2~MK and $3.4\times10^8$~cm$^{-3}$, respectively. The background heating is then gradually decreased to zero over the first two nanoflare cycles, such that the loop is powered entirely by the chromospheric nanoflares thereafter. Runs 1 and 2 follow the evolution of the plasma for 10,000~s of solar time (40 and 20 nanoflares, respectively), and Run 3 follows the plasma for 20,000~s (20 nanoflares). The upper two panels of Figure~\ref{fig2} show the spatially averaged temperature and density as a function of time for Run~1. The red curves in each panel show the corresponding temperature and density from a model run in which the plasma was simply allowed to cool, in order that the ability of the chromospheric nanoflares to sustain coronal temperatures and densities can be assessed. The generalized cooling formula given by \cite{Cargill2014} (Equation~A2) estimates the cooling timescale for the initial conditions as 2.75~hours, which agrees extremely well with the numerical model. The key parameters for each Run are summarized in Table~\ref{table1}.

\begin{figure}
\includegraphics[width=\textwidth]{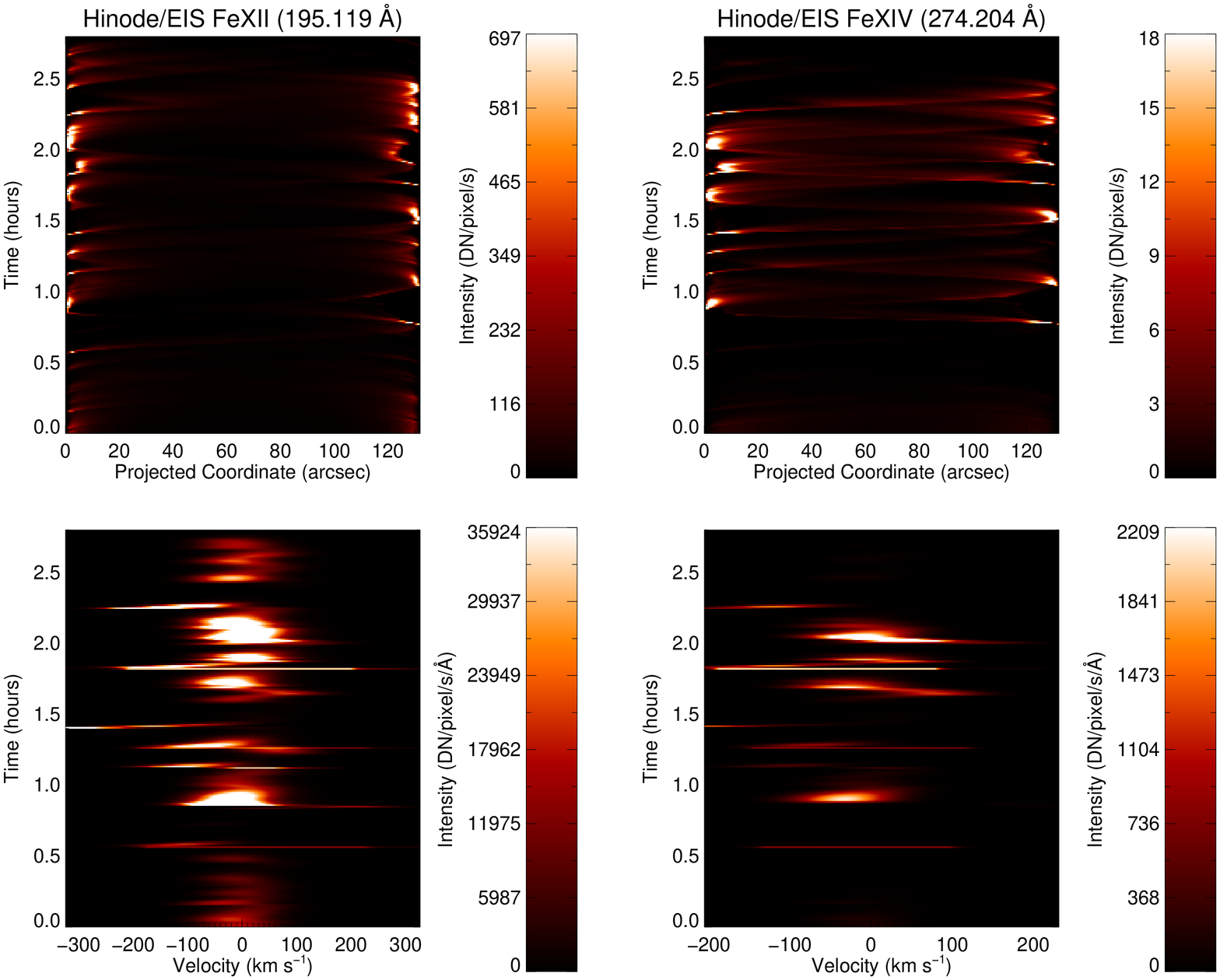}
\includegraphics[width=\textwidth]{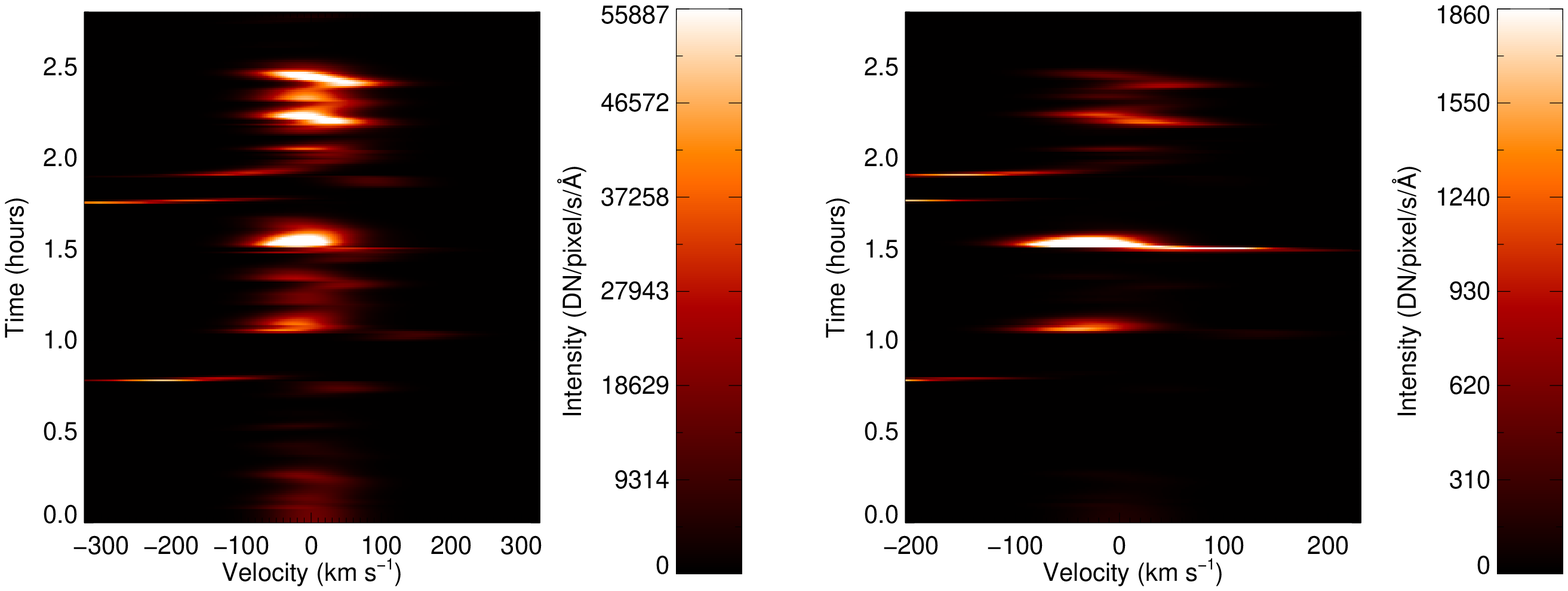}
\caption{Upper panels: the predicted counts, as a function of position along the projected loop and time during Run~1, forward modeled for the Hinode/EIS Fe~XII and XIV lines. Middle panels: the evolution of the spectral line profiles, summed over the first ten pixels, at the left-hand footpoint. Lower panels: the evolution of the spectral line profiles, summed over the last ten pixels, at the right-hand footpoint.}
\label{fig3}
\end{figure}

To provide a quantitative assessment of whether repeating chromospheric nanoflares can sustain the high temperature corona then it is necessary to predict observable quantities from the numerical runs that can be compared directly with real data. Our forward modeling approach follows \cite{Bradshaw2011}, \cite{Reep2013}, and KB14. The emission intensity for the SDO/AIA 171 and 193~\AA~channels is calculated in each grid cell, at one second intervals, from the column emission measure, ion population (the NEI population will be used throughout the remainder of this work), the appropriate wavelength-resolved instrument response, and other quantities (e.g. the element abundance and proton:electron ratio). The predicted count rate is then calculated by projecting the magnetic strand, assumed to be viewed from directly above, onto a one-dimensional detector (aligned with the magnetic field) with pixels the size of SDO/AIA (0.6\arcsec) and adding the contribution from each grid cell to the corresponding detector pixel. Consequently, the line-of-sight through the footpoints is longer than through the apex. The counts are integrated over 10~s intervals to emulate the cadence of AIA. This is equivalent to selecting a loop diameter of 10 pixels. The lower panels of Figure~\ref{fig2} show the results of this procedure for the SDO/AIA 171 and 193~\AA~channels over the full duration of Run~1. The same process is used to calculate the counts in each pixel for Hinode/EIS Fe~XII (195.119~\AA) and Fe~XIV (274.204~\AA) lines, except that a Gaussian line profile is constructed for each grid cell (using the local line-of-sight velocity, the thermal width, and the instrument FWHM) and the detector pixel size is changed to 1.0\arcsec. We adopt a 10~s cadence for consistency with the AIA results, which assumes a favorable alignment of the spectrometer slit with the loop (no rastering required). Figure~\ref{fig3} shows the evolution of the total counts and the chosen spectral line profiles, at the left- and right-hand footpoints, for EIS.

The observable quantities that we calculate from our forward modeled predictions for direct comparison with real data, to determine whether chromospheric nanoflares can power the corona, are the AIA channel/EIS spectral line intensities, the Dopper-shifts of the lines, and their red/blue asymmetries. A full discussion and analysis of our findings follows in Section~\ref{results}.

\section{Results}
\label{results}

\begin{figure}
\includegraphics[width=0.7\textwidth]{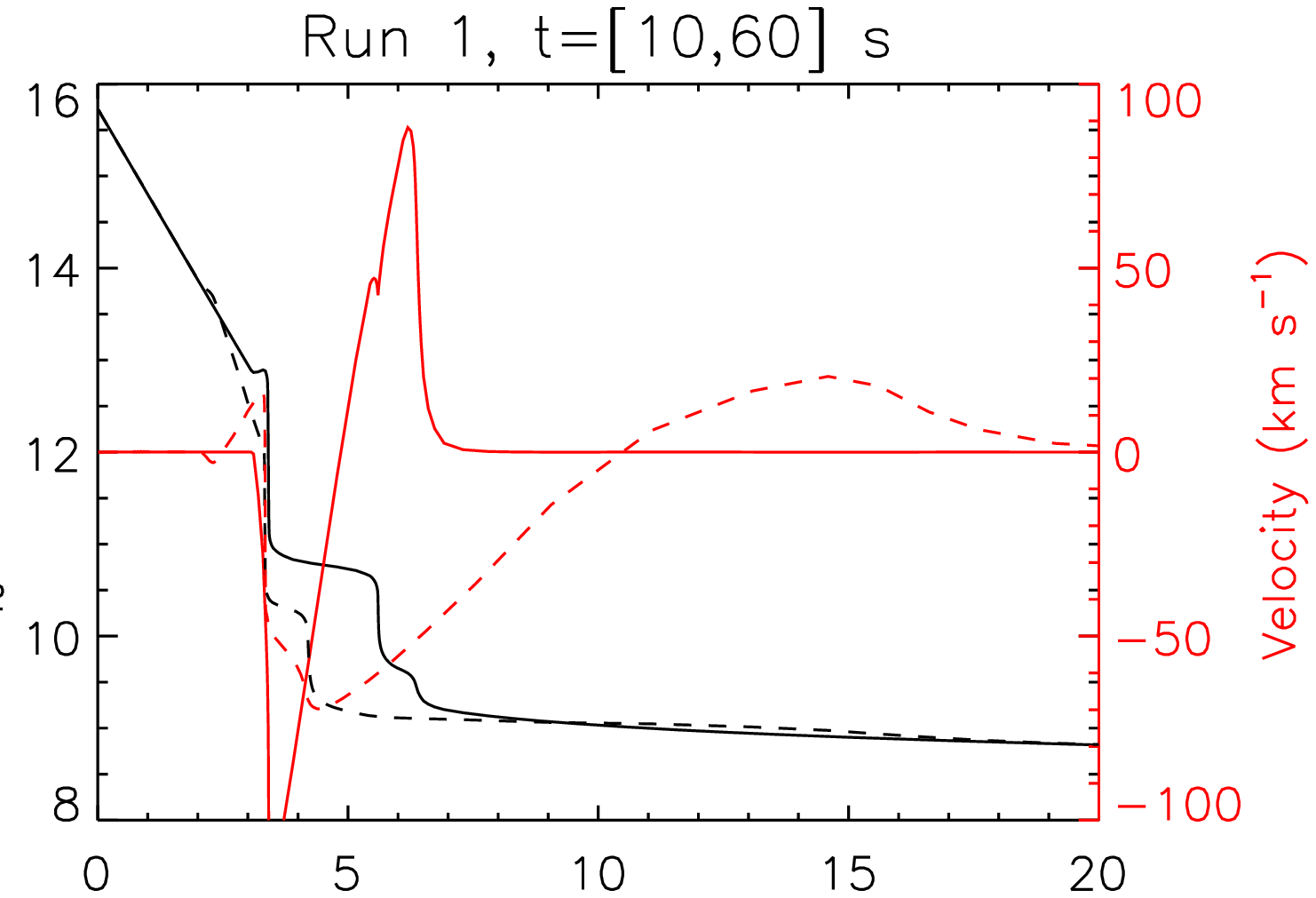}
\includegraphics[width=0.7\textwidth]{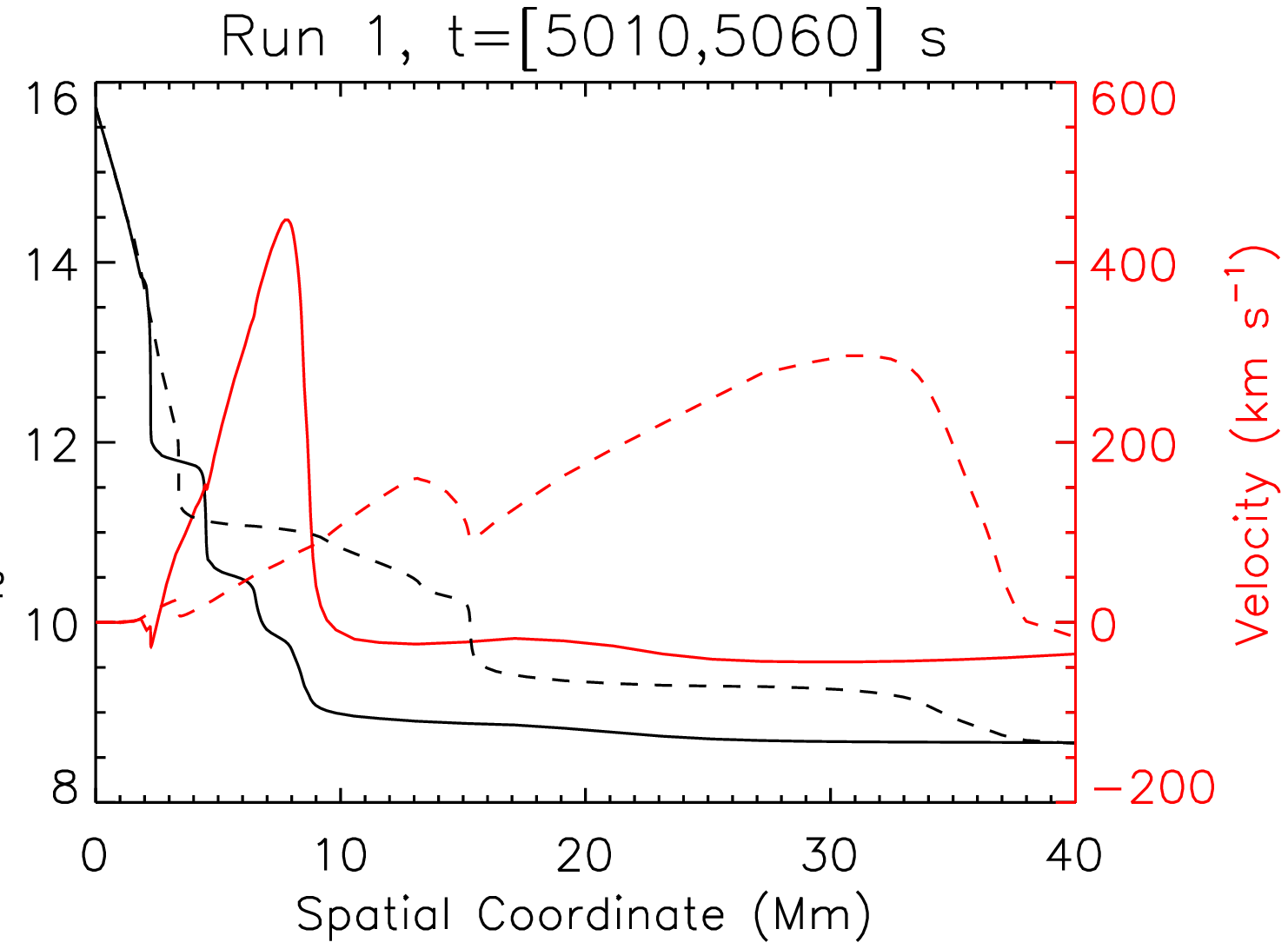}
\caption{The density and velocity profiles 10~s (solid) and 60~s (dashed) after the nanoflares at $t=0$~s (upper panel) and $t=5000$~s (lower panel) during Run~1. The nanoflares are both triggered in the left-hand leg of the loop (note: the full length of the loop is not shown).}
\label{fig4}
\end{figure}

The evolution of each chromospheric nanoflare proceeds similarly to the manner described in KB14. Figure~\ref{fig4} shows the density and velocity profiles after 10~s and 60~s following the nanoflares triggered at $t=0$~s and $t=5000$~s during Run~1. At $t=5000$~s, when the loop is sustained by chromospheric nanoflares alone, the transition region is located approximately 2~Mm deeper in the atmosphere than at $t=0$~s, or about 4.6 scale heights at 0.01~MK. The local pressure is increased by a factor of 200 when the plasma at the base of the transition region is heated to 2~MK and the steep pressure gradient drives a strong expansion that reaches 400~km~s$^{-1}$ at 5010~s, far in excess of the local sound speed ($M\approx10$). Nonetheless, we point out (as in Section~\ref{model}) that the energy available to heat the plasma at greater depths in the chromosphere may not be enough to raise its temperature above 1~MK, given its greater density and heat capacity. That we do not typically observe such strong velocities, even as blue-wing enhancements, in the quiet Sun or in non-flaring active regions lends support to this interpretation. However, to sustain the corona then chromospheric nanoflares must be able to raise the local temperature above at least 1~MK (more in fact, since active region cores are observed closer to $3-4$~MK), with sufficient frequency, wherever they occur.

A plug of relatively high density material is transported upward into the corona by the outflow associated with each nanoflare. If the plug is defined to extend out to the location of maximum upward velocity then its density falls by a factor of $\approx 3.2$ between 10~s and 60~s, and by a factor of $\approx 3.5$ between 5010~s and 5060~s. This is expected due to the increase in volume of the plug as it expands upward and contributes material to the corona. A strong rarefaction follows the material front, leading to strong adiabatic cooling of the expanding plug. This represents a strong physical limitation in terms of the feasibility of chromospheric nanoflares powering the corona. The stronger the heating to raise the chromospheric plasma to increasingly high temperatures then the greater the local pressure, the stronger the pressure gradient driving the expansion and, consequently, the stronger the rarefaction wave and the associated adiabatic cooling of the plug.

Thermal conduction transports energy ahead of the expanding plug. This can heat the lower density material to coronal temperatures. However, the material is quickly overtaken by the plug, which is cooling rapidly. The high temperature material exists for at most the amount of time it takes the plug to fill the loop:   $\sim L/v$, where $L$ is the loop length and $v$ the expansion velocity. Taking the loop length adopted in our study of $157$~Mm and a sustained outflow speed of 400~km~s$^{-1}$, the plug would fill the loop in $\sim 6.5$~minutes. Note that the plug ceases to be a source of thermally conducted energy long before this, due to its expansion cooling. This is reflected by the rapid fluctuations in the coronal average temperature and density profiles (Figure~\ref{fig2}). Thermal conduction heats the plasma ahead of the plug and the arrival of the material front provides additional energy in the form of compression and (some) viscous heating; hence, both the density and the temperature sharply rise. The rarefaction immediately following the plug then drives an expansion; causing a sharp fall in the density and temperature.

A second, much denser plug ($n>10^{10}$~cm$^{-3}$) follows the first plug for the nanoflare at $t=5000$~s. It is driven by the rebound shock due to the restitution of the lower atmosphere in response to the initial expansion, but only reaches a height of $\approx 20$~Mm (less than half the apex altitude). The rapid expansion of the plug cools the plasma to less than 0.02~MK and with an outflow speed of $\sim 100$~km~s$^{-1}$, while maintaining a relatively high density, it could be a viable spicule candidate. A height of 20~Mm is greater than the typical extent of a spicule, but a shorter structure might be consistent with observations and such a mechanism could then fill the loop, albeit with cool plasma (perhaps explaining some transition region structures), if the material front is able to traverse the apex. However, this is beyond the scope of the present work.

Returning to the top two panels in Figure~\ref{fig2}, we see that the repeating chromospheric nanoflares in Run~1 (inter-event period 250~s, or 500~s at the same foot-point) manage to sustain the corona with an average temperature between $1.5-2$~MK and average density between $5-6\times10^8$~cm$^{-3}$ for more than two hours after the background heating falls to zero, and long after the loop would otherwise have cooled and drained (red curves). As a further example, the generalized cooling formula estimates the cooling timescales based on the temperature and density peaks at 0.80~hours and 0.95~hours as 2.4~hours (to 2 s.f.). While the temperature range is reasonable for the corona (particularly the quiet corona) and remains close to 2~MK, the density is not much greater than the density of the hydrostatic initial conditions ($3.5\times10^8$~cm$^{-3}$). It is well-known that the coronal loops clearly observed in Fe~IX, X, and XII emission (TRACE~171 and 195~\AA; AIA~171 and 193~\AA) are significantly over-dense (by factors of $10-1000$) compared to hydrostatic equilibrium \citep{Aschwanden2001,Winebarger2003}. In consequence, we expect the coronae of the magnetic strands subject to repeating chromospheric nanoflares in Run~1 to be faint rather than brightly visible. This is borne out in the lower two panels of Figure~\ref{fig2} where the predicted emission in instrument units (DN~pixel$^{-1}$~s$^{-1}$) for the AIA~171 and 193~\AA~channels is shown as a function of the projected coordinate along the loop and time. Only the loop foot-points are visible in the 171~\AA~channel (the coronal counts are of order 1~DN~pixel$^{-1}$~s$^{-1}$). The foot-point count rate in the 193~\AA~channel is similar to the 171~\AA~ channel, $\sim400$~DN~pixel$^{-1}$~s$^{-1}$, but the count rate in the upper corona is larger ($\leq~70$~DN~pixel$^{-1}$~s$^{-1}$) because the average temperature is close to the formation range of Fe~XII ($T\approx1.6$~MK), which dominates the 193~\AA~channel.  The sensitivity of the 171~\AA~channel peaks at cooler temperatures. \cite{DelZanna2011} measured count rates from a bright region of $\sim7300$ and $\sim4200$~DN~pixel$^{-1}$~s$^{-1}$ for the AIA~171 and 193~\AA~channels, respectively. They also measured count rates from the leg/foot-point of a warm loop in the range of $\sim3500$ and $\sim2300$~DN~pixel$^{-1}$~s$^{-1}$, and background rates of $\sim2300$ and $\sim1500$~DN~pixel$^{-1}$~s$^{-1}$, for the AIA~171 and 193~\AA~channels.

Revisting the question of the loop length, one may argue that shorter loops maintained at the same temperature (2~MK), with commensurately greater density and pressure, should yield a higher count rate. The average temperature and average density in the upper 50\% of the loop are not too far from their initial static equilibrium values (Figure~\ref{fig2}). We therefore cautiously apply a static equilibrium scaling law \citep{Rosner1978} to derive an estimate of the way in which changing the loop length should alter the count rate and the range of lengths that would bring the predicted count rates into agreement with the observed values. Based on the scaling law $T_{max} = c_1 (PL)^{1/3}$ (where $T_{max}$ is the apex temperature, $c_1$ a constant, $P$ the pressure, and $L$ the half-length) and substituting $P=2k_{B}nT_{max}$ (where $k_B$ is the Boltzmann constant and $n$ the number density) it is straightforward to show that $T_{max}^2 \propto nL$. If we then take the ratio of two loops with the same $T_{max}$, but different density and half-length, then $n_1L_1 = n_2L_2$. The scaling law is a statement that their column densities must be equal under this condition. If the count rate scales approximately as the square of the density, then the ratio of the count rates scales as $(L_1/L_2)^2$. Recall the count rate in the 171~\AA~ channel from the bright region observed by \cite{DelZanna2011} was $\sim7300$~DN~pixel$^{-1}$~s$^{-1}$ and the coronal count rate predicted by Run~1 was 1~DN~pixel$^{-1}$~s$^{-1}$). What is the loop half-length required to boost the predicted count rate to the observed value? The ratio of count rates is 7300 and the half-length of our original loop is 78.5~Mm, which yields a required half-length of $78.5/\sqrt{7300} \approx 1$~Mm (1,000~km). Repeating this calculation for the 193~\AA~channel, where the upper limit to the predicted count rate was 70~DN~pixel$^{-1}$~s$^{-1}$, yields a half-length of 10~Mm. Therefore, while adjusting the loop length may potentially help chromospheric nanoflares to explain some of the shortest coronal structures that are observed, it doesn't help to explain the full range of extended bright structures that are visible along their entire lengths. We also note that the loop length has no bearing on some of the other observational tests that they fail, described below.

The repeating chromospheric nanoflares with the properties defined in Run~1 also fall far short of reproducing even the background counts measured by \cite{DelZanna2011} (assuming the background is comprised of unresolved strands). We note that the line-of-sight may intersect several structures, which may account for some of the difference between our predicted intensities and those observed, but there remains the substantial discrepancy that we find the emission falls off very rapidly with increasing height. Furthermore, emission confined to low altitudes also means that Run~1 does not reproduce what is observed at the solar limb.

The top two panels in Figure~\ref{fig3} show the counts predicted for the Fe~XII 195.119~\AA~and Fe~XIV ($T\approx2$~MK) 274.204~\AA~lines detected by Hinode/EIS. The story is similar in that the counts are significantly lower than measured by \cite{DelZanna2011}. The middle and lower panels of Figure~\ref{fig3} show the evolution of the spectral profiles of these emission lines. The middle-left plot shows the profile of the Fe~XII line summed over the left-most ten pixels of the loop and the lower-left plot shows the line profile summed over the right-most ten pixels (the foot-point pixels). The middle-right and lower-right plots show the corresponding Fe~XIV line profiles at the left- and right-hand foot-points. The foot-point Fe~XII emission is generally blue-shifted (by convention $v<0$~km~s$^{-1}$) up to a few tens of km~s$^{-1}$, with the occasional rapid excursion above 100~km~s$^{-1}$. The same is true of the Fe~XIV emission when it appears, although the blue-shifts are somewhat more pronounced. This pattern is actually consistent with observations, since warmer plasma is generally observed to be more strongly blue-shifted \citep[e.g.][]{Hara2008}, indicating an expansion. Nonetheless, the magnitude of the blue-shifts is not, since observations reveal blue-shifts slower than 10~km~s$^{-1}$ in active regions \citep{Doschek2012,Tripathi2012} and the quiet Sun \citep{Chae1998,Peter1999}. However, even the strongest heating events, associated with the fastest upflows, do not provide the corona with much additional material and energy (weaker blue-shifts would naturally provide even less). For example, there is a particularly strong blue-shift of $\sim150$~km~s$^{-1}$ between 2 and 2.5 hours at the left-hand foot-point; the intensity plots show a feature extending out from the foot-point to about 30$\arcsec$ in Fe~XII before its intensity falls away, but which traverses the loop in Fe~XIV. However, the count rate remains low, with fewer than 100~DN~pixel$^{-1}$~s$^{-1}$ in Fe~XII and 10~DN~pixel$^{-1}$~s$^{-1}$ in Fe~XIV, in the upper corona.

\begin{figure}
\includegraphics[width=\textwidth]{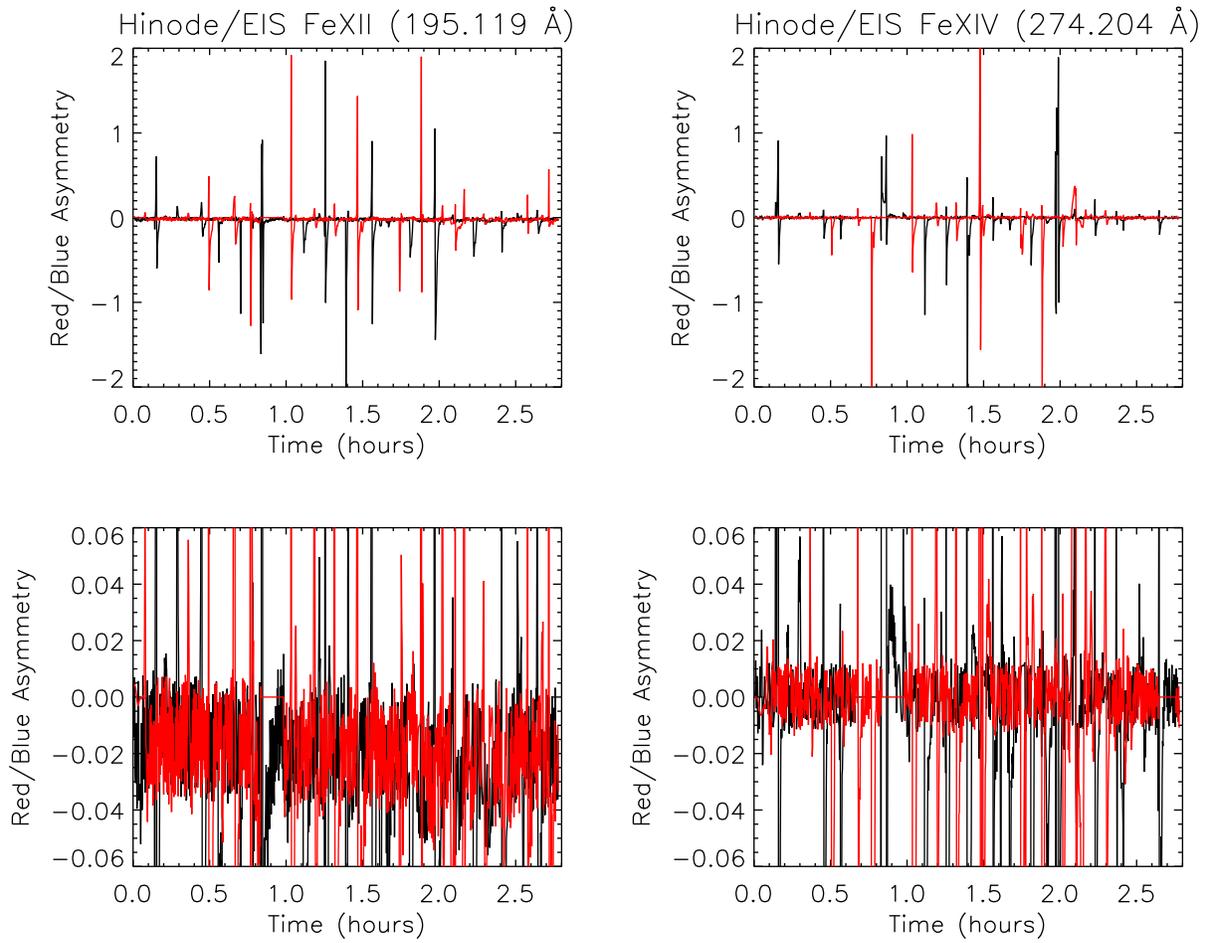}
\caption{The red/blue asymmetries at the left- (black) and right-hand (red) foot-points during Run~1, for the Fe~XII and XIV emission lines. The upper panels show the full range of asymmetries and the lower panels present a restricted range to more clearly show the short-timescale variability.}
\label{fig5}
\end{figure}

The shape of the line profile, in particular the excess of emission in one wing relative to the other, is quantified by the red/blue asymmetry of the line. It is calculated by integrating from -250 to -50~km~s$^{-1}$ in the blue-wing, +250 to +50~km~s$^{-1}$ in the red-wing, and -30 to 30~km~s$^{-1}$ in the line core; the red/blue asymmetry is then defined as the difference between the total red and blue intensities divided by the total core intensity, and it is a positive quantity when the excess emission is in the blue-wing of the line. The rest wavelength of the line is taken to be at $\lambda(I_{max})$ for consistency with observational studies, rather than the true laboratory rest wavelength. Essentially, a correction for the Doppler-shift is applied before the asymmetry is calculated.  Figure~\ref{fig5} shows the red/blue asymmetries at each foot-point during Run~1. 

The asymmetries in the Fe~XII~195.119~\AA~line are generally negative (red-wing excess), with an average of -0.02 and short timescale variability in the range $\sim0.00~\mbox{to}~-0.04$. The reason for the dominant negative asymmetry in this line is a consequence of its formation temperature close to 2~MK: the localized pressure enhancement due to heating the upper 1000~km of the chromosphere drives some Fe~XII emitting plasma surface-ward and some toward the corona; the plasma driven toward the surface is denser due to the gravitational stratification and radiates more strongly, and this red-shifted emission in turn leads to a stronger red-wing excess when the Fe~XII emission is integrated along the line-of-sight. The predicted red-wing excess in the Fe~XII line is inconsistent with observations in active regions, the quiet Sun, and coronal holes, where a blue-wing excess is more common \citep{Hara2008,DePontieu2009,DePontieu2011,McIntosh2009,Tian2011,Doschek2012,Tripathi2013,Patsourakos2014}. There are also regular excursions to values up to $\pm2$, far larger than are measured in observational data. The asymmetries predicted in the Fe~XIV~274.204~\AA~line have an average of zero in the range $\pm0.02$, also with large excursions extending to $\pm2$.

\begin{figure}
\includegraphics[width=\textwidth]{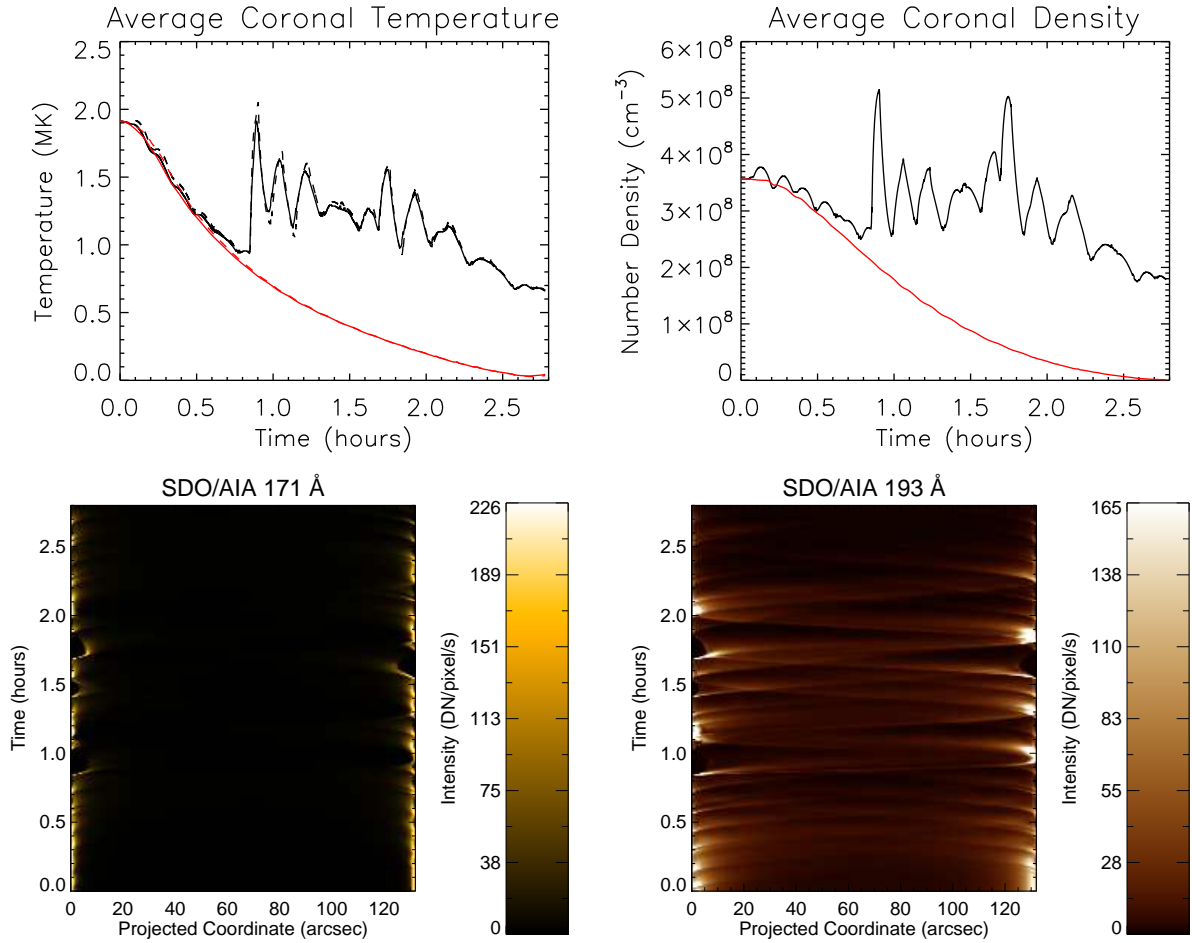}
\caption{Upper panels: the temperature and density (black curves), spatially averaged over the upper 50\% of the loop, as a function of time for Run~2. Lower panels: the predicted counts, as a function of position along the projected loop and time, forward modeled for the SDO/AIA 171 and 193~\AA~channels.}
\label{fig6}
\end{figure}

The upper two panels of Figure~\ref{fig6} show the evolution of the average coronal temperature and density during Run~2. The repeating chromospheric nanoflares (inter-event period 500~s, or 1000~s at the same foot-point) are unable to sustain the corona at a consistent average temperature above 1~MK and the trend over a two-hour period is for cooling and draining to persist. Though the corona ultimately cools into the temperature range within which we might expect AIA~171~\AA~emission to be observed, the density is low enough ($<3\times10^8$~cm$^{-3}$) that the counts are very small compared to the foot-point emission and most of the loop remains invisible in the lower-left panel of Figure~\ref{fig6}. This is not consistent with observations of bright loops in the 171~\AA~channel. The lower-right panel shows the counts predicted in the AIA~193~\AA~channel and while the loop remains visible along most of its length, the counts are reduced by a factor of $\approx2.6$ compared to the higher frequency nanoflares of Run~1, representing an even more marked discrepancy with measured counts from warm loops.

\begin{figure}
\includegraphics[width=\textwidth]{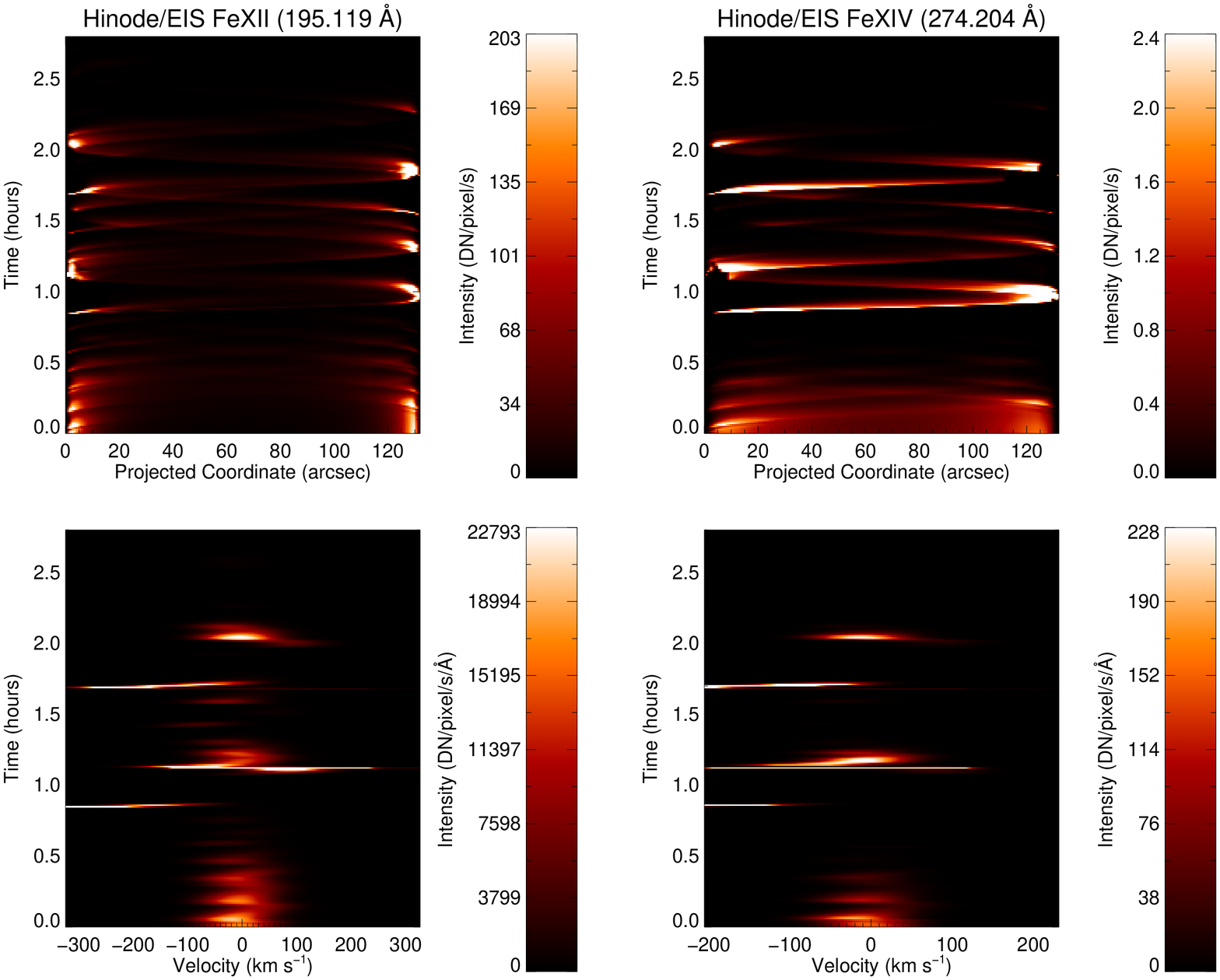}
\includegraphics[width=\textwidth]{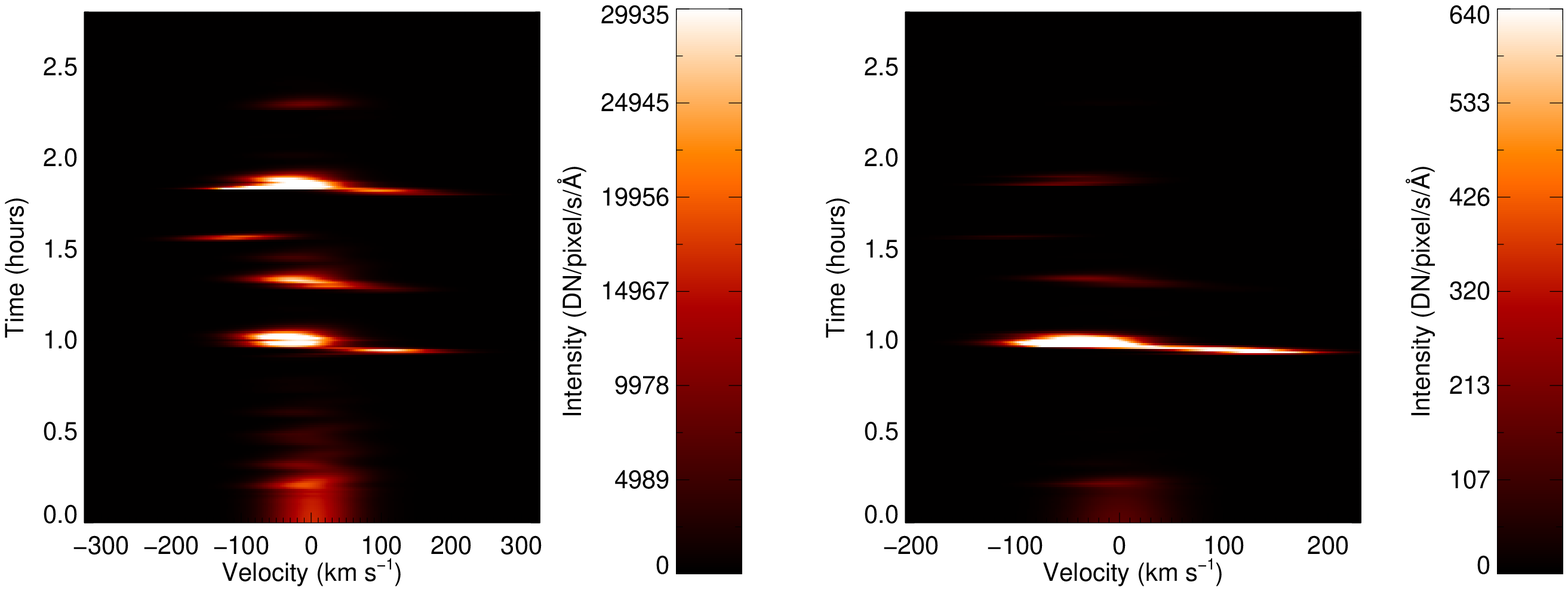}
\caption{Upper panels: the predicted counts, as a function of position along the projected loop and time during Run~2, forward modeled for the Hinode/EIS Fe~XII and XIV lines. Middle panels: the evolution of the spectral line profiles at the left-hand footpoint. Lower panels: the evolution of the spectral line profiles at the right-hand footpoint.}
\label{fig7}
\end{figure}

The upper panels in Figure~\ref{fig7} show the counts predicted for the Fe~XII and Fe~XIV lines during Run~2, and the middle and lower panels show the evolution of the spectral profiles of these emission lines. The intensities are weaker than in Run~1, as expected, and the loop structure appears more intermittently; it does not persist for the periods ($1000-5000$~s) observed. The intermittency is reduced in the AIA channels because there are a large number of lines (several thousand) that fall within each channel and are formed across the range of temperature variability. The cores of the line profile are generally blue-shifted by a few tens of~km~s$^{-1}$, again with the occasional burst above 100~km~s$^{-1}$, but red-shifts above 100~km~s$^{-1}$ are also predicted (particularly at the right-hand foot-point). Such strong down-flows are not typically observed in quiescent active region or in the quiet Sun. The red/blue asymmetries of the Fe~XII and XIV emission lines for Run~2 follow the pattern of Figure~\ref{fig5}, with averages of -0.02 (between -0.04 and 0.00 for Fe~XII) and 0.00 (between $\pm0.02$ for Fe~XIV). Stronger excursions also arise, but are generally limited to $\pm1$ in this case.

\begin{figure}
\includegraphics[width=\textwidth]{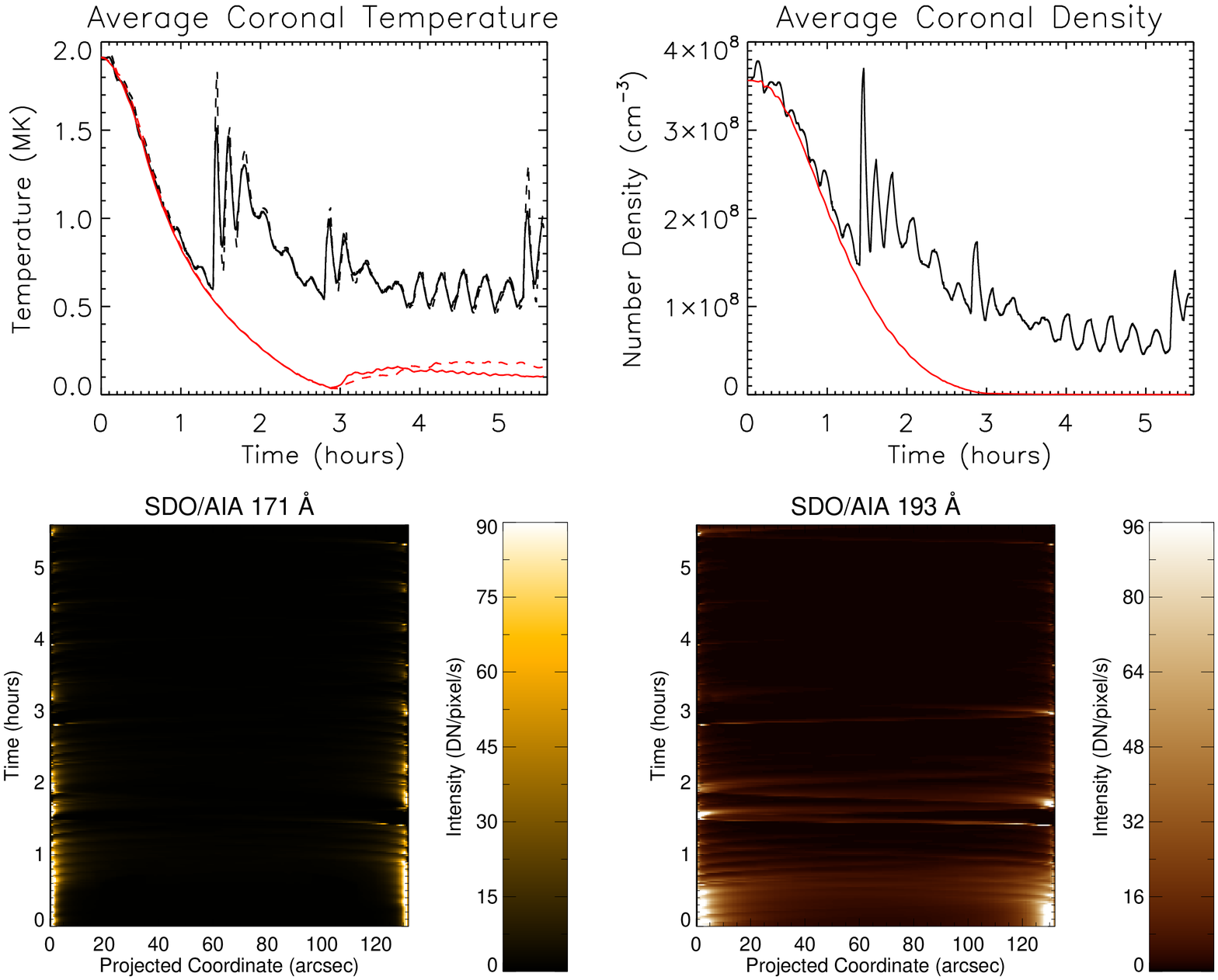}
\caption{Upper panels: the temperature and density (black curves), spatially averaged over the upper 50\% of the loop, as a function of time for Run~3. Lower panels: the predicted counts, as a function of position along the projected loop and time, forward modeled for the SDO/AIA 171 and 193~\AA~channels.}
\label{fig8}
\end{figure}

\begin{figure}
\includegraphics[width=\textwidth]{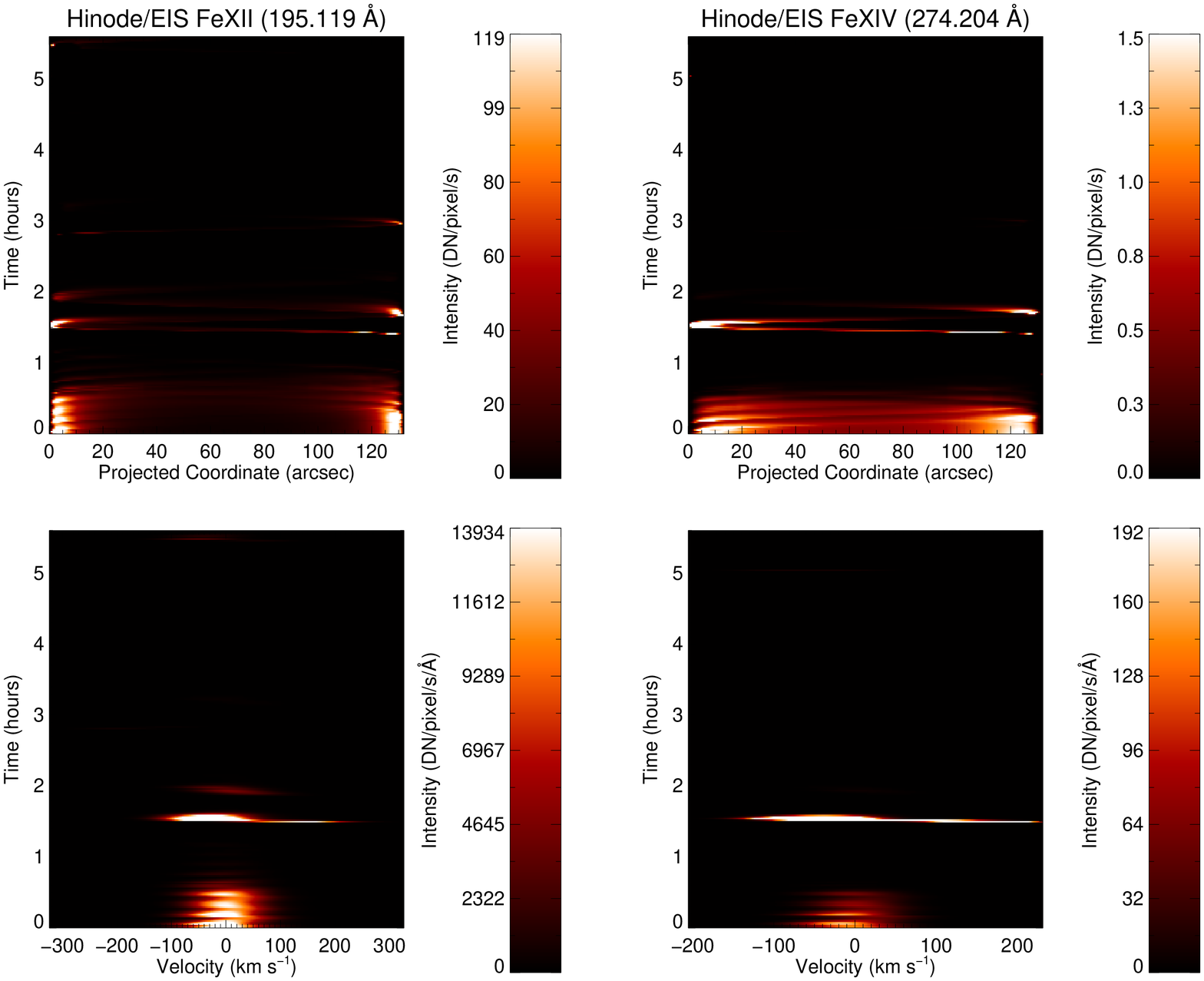}
\includegraphics[width=\textwidth]{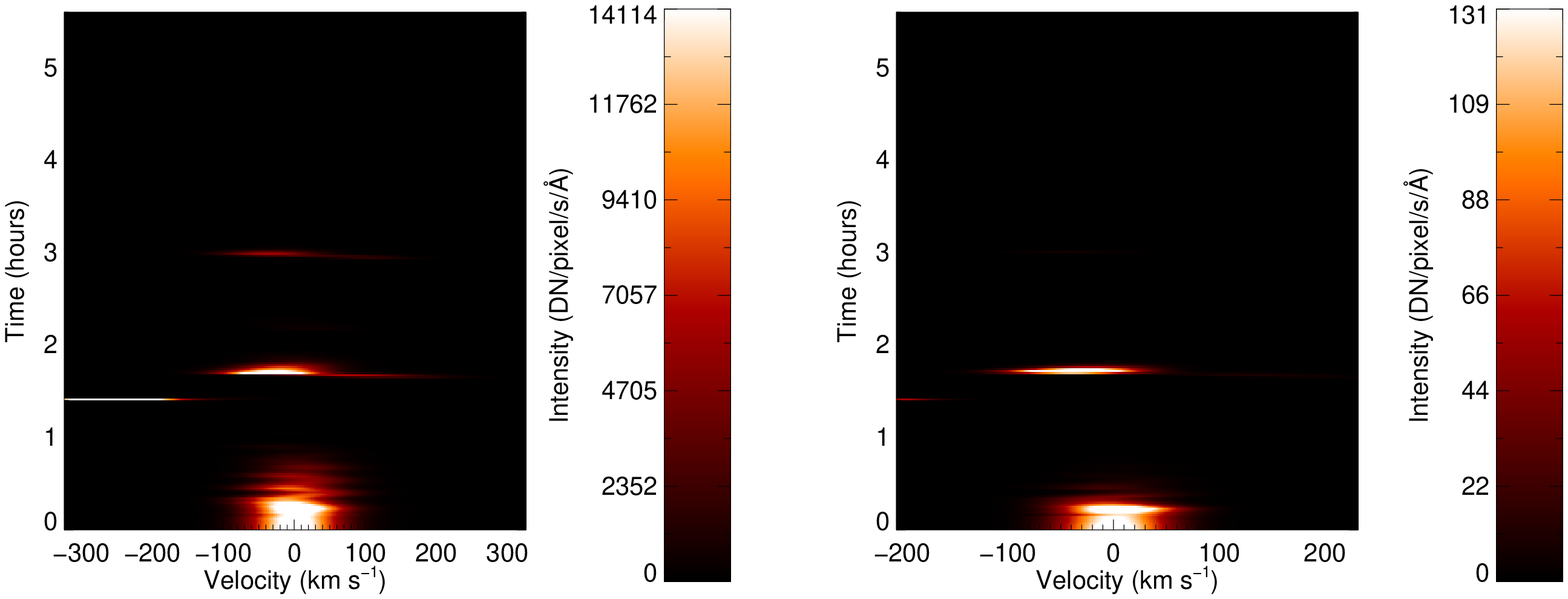}
\caption{Upper panels: the predicted counts, as a function of position along the projected loop and time during Run~3, forward modeled for the Hinode/EIS Fe~XII and XIV lines. Middle panels: the evolution of the spectral line profiles at the left-hand footpoint. Lower panels: the evolution of the spectral line profiles at the right-hand footpoint.}
\label{fig9}
\end{figure}

Figures~\ref{fig8} and \ref{fig9} show the plots corresponding to Figures~\ref{fig6} and \ref{fig7} for Run~3 (inter-event period 1000~s, or 2000~s at the same foot-point). The coronal part of the loop cools to an average temperature of $\approx0.6$~MK and drains to a density of about $6\times10^7$~cm$^{-3}$ over the $\sim5$~hour period of the numerical experiment. The intensities are weaker still, such that the full extent of the structure is never observed in the AIA~171~\AA~channel and it fades from view after $2.5-3$~hours in the 193~\AA~channel (even the foot-point emission is weak in both channels after this time). The intensities are a factor of 2.5 and 1.7 weaker than Run~2 in the 171 and 193~\AA~channels, respectively. This behavior is also seen in the predicted Fe~XII and XIV emission, except that the loop structure disappears even sooner in the hotter line, accompanied by weaker blue-shifts (the line cores are close to 0~km~s$^{-1}$ early on, especially at the right-hand foot-point) than the previous Runs and the occasional, strong blue-/red-shift above $\sim50$~km~s$^{-1}$. The red asymmetry in the Fe~XII line gradually decreases over a 3~hour period from -0.02 to -0.01, where it remains for the remainder of Run~3, at both foot-points, and the stronger deviations are generally less than $\pm1$. The Fe~XIV line asymmetry maintains an average of $\approx$~0 with less frequent deviations typically below $\pm0.5$.

\section{Summary and Conclusions}
\label{SandC}

We have performed several numerical experiments in which the response of the solar corona to a train of repeating chromospheric nanoflares is investigated. The inter-event period is varied between 250 and 1000~s, with nanoflares triggered at alternating foot-points (500 to 2000~s at a single foot-point). The purpose of this investigation is to extend the study of KB14, in which the plasma on a single magnetic strand is energized only once in a 5000~s period (essentially equivalent to a 5000~s inter-event period), by determining whether repeating chromospheric nanoflares can lead to an accumulation of warm material and produce fully developed coronal strands with the observed properties. To resolve this question we have compared predicted temperatures, densities, intensities (narrow-band channel and individual spectral lines), Doppler-shifts, and line profiles (via their red/blue asymmetry) with real observations. Only a consistent match across all observables is sufficient to pin down the requisite properties of the repeating nanoflares.

Beginning with an inter-event period of 250~s (Run~1), at the lower end of the range found to be consistent with observed emission measure slopes \citep{Cargill2014}, we find that coronal temperatures can be sustained for several hours after the background heating has been ramped down to zero, by which time the loop would have fully cooled with no additional energy input, but that the observed over-densities (factors of $10-1000$ greater than hydrostatic) are not reproduced. Consequently, the predicted intensities in the AIA~171 and 193~\AA~channels are much weaker than observed in warm coronal structures. No fully developed loops are observed in the 171~\AA~channel and the count rate in the 193~\AA~channel is low. The numerical experiments therefore cannot account for the high altitude loops observed at the solar limb. The predicted intensities of two coronal emission lines (Fe~XII at 195.119~\AA~and Fe~XIV at 274.204~\AA) observed by EIS are similarly weaker than observed. In addition, the Doppler-shifts are generally too large (tens of km~s$^{-1}$), with large excursions of the line core to bulk upflows exceeding 100~km~s$^{-1}$, which are not typically observed in non-flaring active regions and in the quiet Sun. The magnitudes of the red/blue asymmetries are comparable to those observed (few percent), but the enhancements are usually in the red wing of Fe~XII, whereas blue wing enhancements are more typically observed. Also, there are regular deviations to much larger values ($\pm2$) that, again, are not seen in observational data. Considering chromospheric nanoflares on a large number of sub-resolution strands would not help to resolve these discrepancies: far too many strands would be required for complete structures to form in the 171~\AA~channel and, regardless, simply summing strands would not produce blue-wing asymmetries from individual strands that exhibit red-wing asymmetries in Fe~XII emission.

Overall, our findings demonstrate that even the most energetically effective heating events located in the chromosphere fail to produce and sustain a corona with the desired properties, due to the rapid cooling associated with a fast expansion. Increasing the inter-event period to 1000~s (Runs~2 and 3) simply exacerbates the discrepancies between the predicted and measured observables. The average temperature cannot be sustained at typical coronal values, the densities are lower, and the intensities are commensurately weaker. The predicted intensity falls by a factor in the region of $2-2.5$ as the inter-event period doubles, which is consistent with the estimate given in the Discussion (Section~3) of KB14. The Doppler-shifts and the red/blue asymmetries are less sensitive to the inter-event period; there is a gradual trend toward lower values at later times during Run~3, but still exhibiting occasional departures to extreme values well beyond the observed range.

The initial conditions are approximately one order of magnitude denser in the present work, compared to the initial conditions adopted by KB14. As also discussed in Section~3 of KB14, increasing the density in the chromosphere necessarily increases the coronal density but, nonetheless, the upper part of the loop still does not become sufficiently dense when powered by chromospheric nanoflares for the predicted intensities to match those observed in the warm corona. Clearly, increasing the inter-event period yet further would not resolve this difficulty. Decreasing the inter-event period below 250~s may help, but this begins to approach quasi-steady heating, which we know is inconsistent with many observations (e.g., over-dense loops and shallow emission measure slopes).

The way we impose heating in our simulations optimizes the chances of reproducing the observed corona. We raise the temperature of the upper chromosphere (as defined by temperature) to 2 MK. This takes a variable amount of energy that depends on the density of the upper chromosphere at the time the nanoflare is imposed, and it occurs at a variable height, since the chromosphere moves up and down in response to changing coronal pressure. One could argue that this is artificial. Perhaps nanoflares have energies that are independent of plasma density. Perhaps they occur at locations that are determined by the magnetic field rather than by the height of the chromosphere. If this were the case, the plasma would often be heated to temperatures much different from the optimal 2~MK. Some proportion of nanoflares would produce only sub-coronal temperatures, and others would produce extremely high temperatures. If the chromosphere has to move up in order for nanoflares to become more energetically effective then the coronal pressure must fall, which means that it must cool and drain a certain amount before it can be re-energized. This may place a fundamental limit on the amount of heated material that chromospheric nanoflares could provide to the corona.

We conclude that chromospheric nanoflares, repeating on any timescale, are not primarily responsible for powering the observed coronal emission. The most energetically effective coronal heating must take place above the chromosphere. Nonetheless, this does leave open the possibility that chromospheric nanoflares may provide a substantial energy input (and perhaps even the dominant component) to the chromosphere.

\acknowledgements
This work was supported by the NASA Supporting Research and Technology Program. The authors also benefited from participating in the team hosted by the International Space Science Institute, Bern, on Using Observables to Settle the Question of Steady versus Impulsive Coronal Heating, led by SJB and Helen Mason. We also thank the referee for suggestions which helped us to clarify several points.


\end{document}